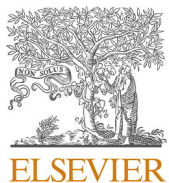
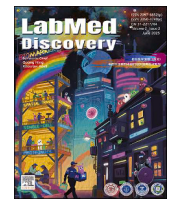

Research Article

# A pilot cohort study of a microfluidic-based point-of-care bilirubin measurement system

Jean Pierre Ndabakuranye [1,2], Inge W.G. Last [1,3], Kay Weng Choy [4], Peter Thurgood [1], Jason C. Steel [5], Genia Burchall [5], Stella Stylianou [6], Khashayar Khoshmanesh [1], Arman Ahnood [1,*]

[1] *School of Engineering, RMIT University, Victoria 3000, Australia*
[2] *Department of Electrical and Electronic Engineering, University of Melbourne, Victoria 3010, Australia*
[3] *Faculty of Science and Engineering, University of Groningen, AG Groningen 9747, the Netherlands*
[4] *Northern Health, Northern Pathology Victoria, Victoria 3076, Australia*
[5] *School of Health, Medical and Applied Sciences, Central Queensland University, Rockhampton 4702, Australia*
[6] *School of Science, RMIT University, Melbourne 3000, Australia*

## HIGHLIGHTS

- A novel point-of-care device for measuring bilirubin concentrations.
- Device uses an optoelectronic sensing module and microfluidic cartridge.
- High clinically accuracy coefficient of determination compared to standard method.
- Strong diagnostic capabilities, achieving a sensitivity of 90% and specificity of 97%.

ABSTRACT

*Objective:* The concentration of bilirubin in blood or serum is useful for assessing liver function as well as monitoring treatment. This study evaluates the clinical performance of a novel point-of-care (PoC) device for the detection of bilirubin in serum. The PoC device incorporates an integrated miniature optoelectronic sensing module and a microfluidic test cartridge.
*Methods:* Patients' serum total bilirubin concentrations, ranging from 2 μmol/L to 480 μmol/L, were measured using the PoC device and the standard laboratory method ($n$=20). Bland-Altman analysis and regression analysis using Passing-Bablok method were used to benchmark the PoC device against the standard laboratory measurements. The diagnostic capability of the PoC device in categorising the serum samples within clinically relevant bilirubin concentration thresholds of 200, 300, and 450 μmol/L was assessed using receiver operating characteristic (ROC) analysis.
*Results:* The mean difference between the PoC device and the standard laboratory method was −5.6 μmol/L, with a 95% confidence interval (*CI*) of −45.1 μmol/L to 33.9 μmol/L. The coefficient of determination ($R^2$) was 0.986. The PoC device achieved a detection sensitivity of 90% and specificity of 97% in categorising bilirubin concentrations within bands used in clinical decision-making.
*Conclusions:* This study demonstrates that the proposed PoC device is capable of measuring bilirubin levels in patient samples with clinically acceptable accuracy.

## 1. Introduction

Serum bilirubin is a key biomarker used to assess liver function. Total bilirubin concentration is correlated with liver physiology and pathology.[1] It is widely used in tracking liver health and treatment effectiveness, such as in neonatal hyperbilirubinemia. Bilirubin concentration can be measured using optical and chemical techniques in clinical settings.[2] However, due to their complexity, cost, and space requirements, existing laboratory techniques are not suitable for point-of-care (PoC) applications.[2]

Several studies have demonstrated the viability of PoC devices for bilirubin measurement in blood samples.[3–6] These devices utilise the distinct optical absorption signature of bilirubin, operating by measuring light absorption at 2 wavelengths, typically 465 nm and 570 nm. Another study reported using wavelengths of 450 nm and 580 nm, with an additional 660 nm for the measurement of the sample holder.[7] Our earlier work reported a miniaturised and scalable PoC device for bilirubin measurement in whole blood using the two-wavelength technique at 455 nm and 530 nm.[8–11] Unlike other PoC devices, ours includes a fully integrated miniatured optical sensing module and microfluidic test cartridge. The device is economical for scalable adoption and highly portable due to its millimetre-sized sensing element. Our previous studies focused on measuring bilirubin concentration using whole blood samples from animal models spiked with exogenous bilirubin.[9,12]

This work evaluates the efficacy of our PoC device in measuring bilirubin concentration in human patient samples with varying endogenous bilirubin levels. Test cartridges with embedded microfluidic channels integrated with the optical sensor ensured a consistent optical path whilst using small sample volumes. Although we validated the setup for direct measurement of bilirubin in whole blood samples in our earlier studies,[12,13] this study focuses on using blood serum to minimise the potential impact of haemoglobin variability and its interference in the measurement. This approach was taken given the small scale of this pilot cohort study. The diagnostic accuracy of the PoC device was compared to the standard clinical laboratory method known as the diazo method. The results were analyzed to determine the statistical association between measurements by the standard laboratory assay and the PoC device. The accuracy of our PoC device in detecting bilirubin levels through receiver operating characteristic (ROC) analyses against the reference values is assessed.

## 2. Materials and methods

### 2.1. Materials

#### 2.1.1. Sample collection

The patients were positioned comfortably with their veins (typically antecubital) accessible. After applying a tourniquet and cleaning the puncture site, venepuncture was performed, followed by blood collection into a vacutainer tube designed for serum separation. Samples were centrifuged at 2,000 g to separate the serum from blood cells. Approximately 100 μL of serum was transferred to a multi-well sample tray using a sterile pipette, ensuring each specimen was uniquely identified. Measurements were carried out at Northern Pathology Victoria (Northern Health, Epping, Victoria, Australia) with an ethics approval number Northern Health NR-23-04.

#### 2.1.2. Sample characteristics

The specimens were residual samples from patients whose blood was collected for routine clinical reasons, with bilirubin testing requested. Due to the potential for assay interference caused by specimen haemolysis, only serum specimens with haemoglobin concentrations (haemolysis level) below 500 mg/dL were selected. Samples were analyzed using the PoC method and the Alinity c simultaneously to avoid any confounding factors, such as sample deterioration due to light exposure.

### 2.2. Methods

#### 2.2.1. Reference laboratory measurement methods

The reference laboratory bilirubin values were based on the standard laboratory method using Abbott Alinity c system (Abbott Laboratories, IL, USA). The Abbott Alinity c Total Bilirubin assay has been standardized against the Doumas method/National Institute of Standards and Technology Standard Reference Material (NIST SRM 916) based on Alinity c Bilirubin Calibrator Kit (REF 0896101). The measurement principle used here is based on the coupling of Total bilirubin with a diazo reagent to form azobilirubin. Absorbance of azobilirubin is measured at 548 nm using Alinity c Total Biliruin Reagent Kit (REF 04V5121). The RCPAQAP analytical performance specifications for the Neonatal Bilirubin program (±8 μmol/L for results up to 80 μmol/L and ±10% for results greater than 80 μmol/L) are based on allowable imprecision of assays and have been used to comment on the performance of individual platforms.[14]

#### 2.2.2. Point-of-care measurement method

The PoC device measurement involved placing the blood serum samples in a test cartridge containing the microfluidic channel. The test cartridge was inserted into the device on top of the sensor system. Data was recorded on a PoC for further analysis. An optoelectronic chip from earlier works was used at the heart of the PoC device,[15] with modifications to include an enclosure for the new test cartridge.

#### 2.2.3. Fabrication of microfluidic system

The microfluidic test cartridge was constructed using a method scalable, consisting of a glass slide and Polydimethylsiloxane (PDMS) slab. Master structures were designed using SolidWorks (Dassault Systèmes) and patterned onto a 200 μm thick layer of SU-8 3050 photoresist (Kayaku Advanced Materials Inc.) that was spun onto a 100 mm diameter crystalline silicon wafer, using a maskless aligner (MLA 150, Heidelberg Instruments). PDMS structures were fabricated by mixing the base and curing agent to a ratio of 1–10 (by weight, Sylgard 184, Dow) and pouring over the master structures to a height of 4 mm. The PDMS was cured in a levelled oven at 130°C for 20 min, then peeled from the master moulds and cut to 75 mm×25 mm using a scalpel. Inlet/outlet ports were created using a 5 mm biopsy punch (Harris Uni-core). Finally, the PDMS slabs containing microfluidic structures were permanently bonded to glass microscope slides (75 mm×25 mm×1 mm) using a plasma cleaner (PDC-002, Harrick Plasma).

#### 2.2.4. Sensing approach

The absorption spectrum of bilirubin shows an absorption peak in the blue region, centred at 455 nm, with minimal absorption in the green and red regions. Given bilirubin's distinct absorption peak at 455 nm and negligible absorption at 530 nm, these wavelengths are ideal for a dual-wavelength ratiometric concentration measurement approach. The selection of the 2 wavelengths has been extensively described in our earlier work.[2] Eq. (1) shows the relation between the sensor light intensities at blue ($I_{blue}$) and green ($I_{green}$) and the diagnostic parameter ($\rho$). In our measurement configuration, $I_{blue}$ and $I_{green}$ are the reflected light intensities measured by the sensor, inversely correlated to the light absorbed by the sample. $I_{blue}$ is measured using an LED with a wavelength of 455 nm, and $I_{green}$ is measured using an LED with a wavelength of 530 nm.

$$\rho = \frac{log(I_{blue})}{log(I_{green})} \quad (1)$$

Eq. (2) is used to calculate the bilirubin concentration ([BR]) using $\rho$. Baseline constant values related to the LED emission intensities, optical properties of the setup, and blood chromophores' spectral signatures are represented as constants A, B and K in Eq. (2).

$$[BR] = (A - B) \times e^{K \times \rho} + B \quad (2)$$

In earlier works, a linear relation between [BR] and $\rho$ was identified. However, various nonlinear mechanisms lead to deviation from the linear relation we had reported in our earlier work.[2,8] This necessitated the use of Eq. (2) in this work.

#### 2.2.5. Integrated sensor module

An integrated miniature sensor (MAX86916, Analog Devices Inc.) was used, with dimensions of 3.5 mm×7 mm×1.5 mm. The sensor chip incorporates multiple LEDs, including green and blue LEDs and a silicon photodiode for measuring the reflected light intensities. The sensor includes all the electronic circuitry required for low noise light intensity measurement and LED operation. Sensor readouts are digitised and transmitted to an external microcontroller unit and subsequent PoC for further analyses.

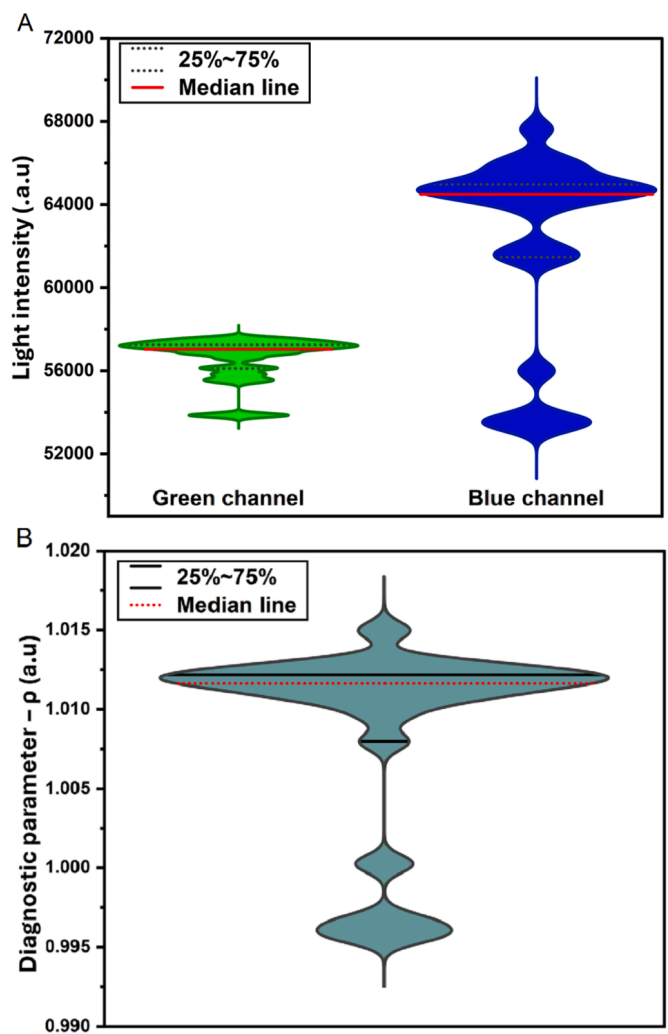

**Fig. 1.** Distribution of PoC device's readout data. Violin plots showing (A) the distribution of the sensor's blue and green channel readouts and (B) the extracted diagnostic parameter ($\rho$) calculated using Eq. (1) for all of the patient population
PoC: point-of-care

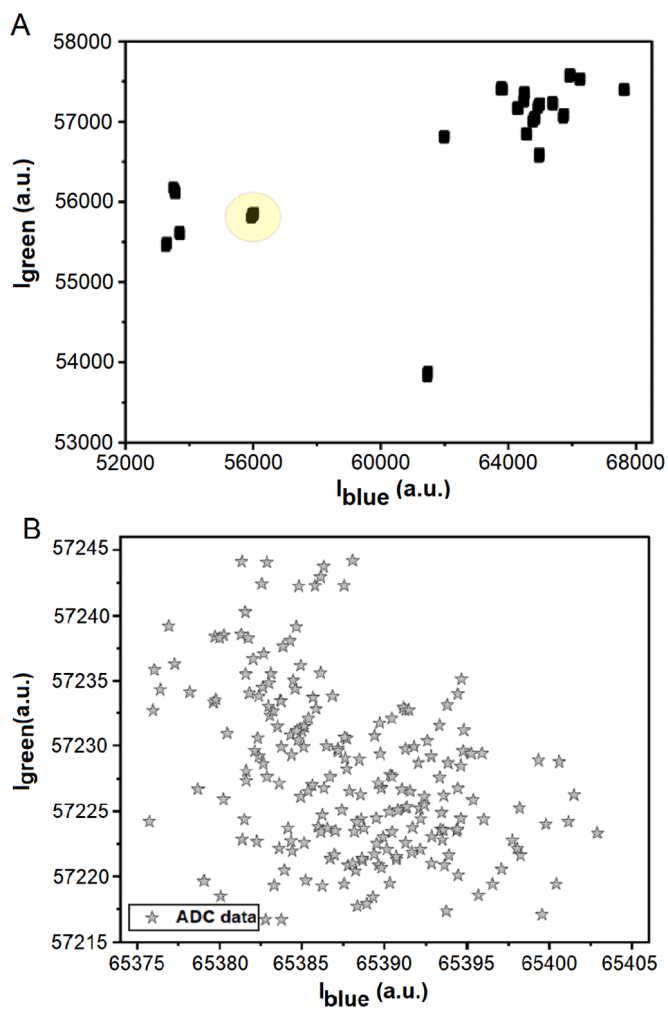

**Fig. 2.** Correlation between the blue and green sensor channel outputs measured by the PoC device; (A) All the samples collected in this study; (B) Single patient highlighted by the yellow circle in (A).
PoC: point-of-care

*2.2.6. Statistical analysis*

Data were analyzed by GraphPad Prism version 10.2.2 (GraphPad Software, Boston, USA) and OriginPro 2024 (OriginLab Corporation, Massachusetts, USA). The correlation between the PoC device readout and bilirubin concentrations was assessed using the Spearman rank correlation method. Bland-Altman regression analysis determined the agreement between bilirubin concentrations measured by the PoC device and the standard laboratory method. The diagnostic efficiency of the PoC system relative to the standard method was evaluated through a ROC curve and a confusion matrix.

### 3. Results

*3.1. Optical data presentation and curation*

Figs. 1A and 2B depict violin plots of the mean and distribution of all sensor readout data points for the blue and green channels. The sensor readout data is proportional to the intensity of the light reflected by the sample in the microfluidic channel. The intensity of the light reflected back to the sample is quantitatively correlated to the inverse of absorbed light intensity. In other words, higher sample light absorption results in lower light reflection. As shown in Fig. 1A, The sensor's blue channel readout has a larger mean value and wider distribution than the green channel due to baseline intensity differences. The green channel readout exhibits a narrow distribution of data points close to the mean, with 4 additional distinct clusters. A similar trend of distinct clusters can be observed in the violin plot for the blue channel. To better understand the reason for this, the ratio of blue to green channels was calculated and presented as a violin plot, showing similar trends.

The scatter plot of blue channel readout vs. green channel readout for all 20 patient samples, indicating a weak correlation between the blue channel and green channel's readout (i.e. lower blue channel recordings typically are weakly predicted by lower green channel recordings and vice versa), suggesting the lack of utility in using a single channel to extract the bilirubin concentration (Fig. 2A). The stability of the recording over the course of a single measurement is an important consideration. The scatter plot of blue channel versus green channel recordings for a single patient sample during 1 measurement cycle of 200 s (Fig. 2B). The green and blue channel readouts do not exhibit a clear correlation ($R^2$ of 0.006,4 using linear regression model), which points to the stability of the recording over a single measurement cycle of 200 s.

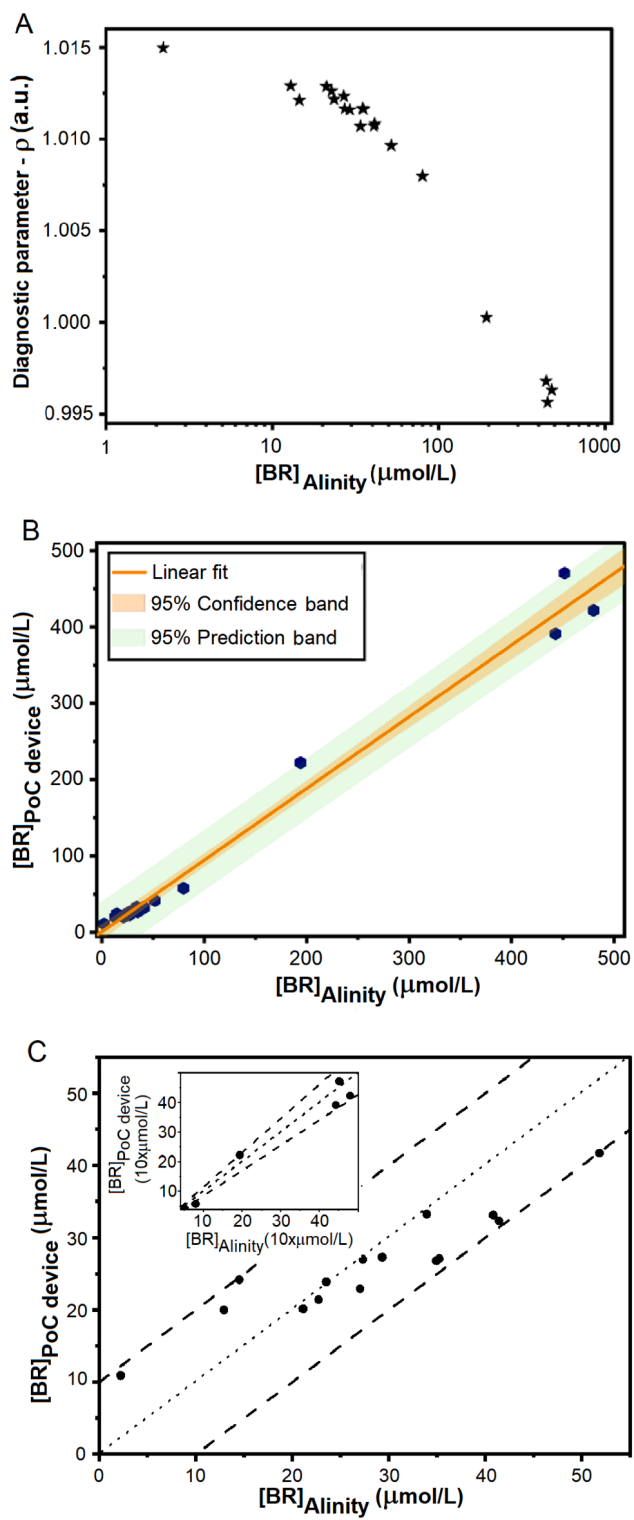

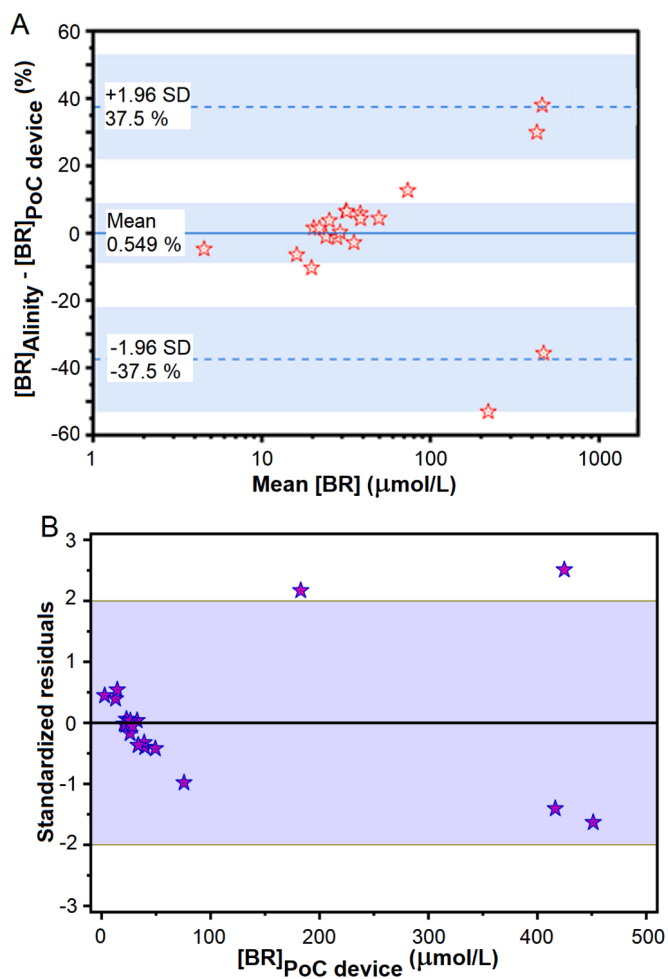

**Fig. 3.** Relationship between the PoC device's measurement and bilirubin concentration as measured using the clinical system; (A) Relationship between the diagnostic parameter ($\rho$) for the point-of-care device and the bilirubin concentration measured using the standard clinical system ([BR]$_{Alinity}$). Eq. (2) is used to convert the diagnostic parameter to bilirubin concentration from the PoC device ([BR]$_{PoC\ device}$); (B) Regression plot showing [BR]$_{PoC\ device}$ against [BR]$_{Alinity}$; The 95% confidence and prediction bands are highlighted; (C) Scatter plot showing [BR]$_{PoC\ device}$ against [BR]$_{Alinity}$ in the 0–50 μmol/L range. The lines indicate the ±10 μmol/L acceptance criteria band; Inset: 50–500 μmol/L range with ±15% acceptance criteria band
PoC: point-of-care

**Fig. 4.** Assessment of the agreement and error in PoC device measurement and the clinical system; (A) Bland–Altman graph (difference) showing the agreement between the bilirubin concentration measured by the PoC and laboratory method; (B) Standardized residual plot for a comparative diagnostic study between measurements by the clinical system and the PoC device
PoC: point-of-care

### 3.2. Correlation and regression analyses

The scatter plot in Fig. 3A shows the $\rho$ measured using the PoC system as a function of the total bilirubin measurement using the Abbott Alinity (diazo method).[16] Spearman rank correlation analysis yielded a coefficient of 0.954 ($P<0.000,1$), indicating significant correlation. Eq. (2) describes the relationship between $\rho$ and the PoC bilirubin concentration. Bland-Altman analysis showed a mean difference of −5.6 μmol/L between PoC device ([BR]$_{PoC\ device}$) and standard laboratory method ([BR]$_{Alinity}$), with a 95% *CI* between −45.1 μmol/L and 33.9 μmol/L. The results of regression analysis between [BR]$_{PoC\ device}$ and [BR]$_{Alinity}$ using the Passing-Bablok method are illustrated in Fig. 3B where a slope of 0.94±0.02 of an $R^2$ of 0.986 was obtained. Fig. 3C shows the measured values along with acceptance criteria of ±10 μmol/L for concentrations of less than 50 μmol/L. The insert plot in Fig. 3C shows the ±15% acceptance criteria for concentrations of above 50 μmol/L. All measured values in study fall within these criteria.

The Bland-Altman plot, indicating no systematic bias in the bilirubin measured using the PoC system when compared to the lab baseline values (Fig. 4A). About 70% of the measurements being within the ±10 % band and 80% within ±20 % band. The residual analysis plot with the mean difference and the standard deviation compared to the Abbott Alinity measurement(Fig. 4B). Byond 2 outlier data points, the remaining measurements are within a standardized residual band. It should be noted that the $[BR]_{Alinity}$ and $[BR]_{PoC\ device}$ results do not exhibit normal distribution, and therefore the results of Bland-Altman should be considered with caution. Nevertheless, the variability does not appear to be systematic and may be attributed to inherent variability in the sample loading and device usage procedures.

**Table 1**
Wilcoxon signed-rank test analysis

|  | Number | Mean rank | Sum of ranks |
| --- | --- | --- | --- |
| Negative ranks [a] | 6 | 11.67 | 70.00 |
| Positive ranks [b] | 14 | 10.00 | 140.00 |
| Ties [c] | 0 |  |  |
| Total | 20 |  |  |

[a] $[BR]_{PoC\ device} > [BR]_{Alinity}$
[b] $[BR]_{PoC\ device} < [BR]_{Alinity}$
[c] $[BR]_{PoC\ device} = [BR]_{Alinity}$

The Wilcoxon signed-rank test, a non-parametric equivalent to the paired samples *t*-test, was used to check for a significant change in the clinical performance of the PoC device in the detection of bilirubin in serum. This non-parametric test was selected due to the small sample size (pilot study), which made testing and assuming normality difficult. As the sample size was small and there were no ties present in the ranks of the data, an asymptotic significance two-tailed *P*-value was calculated as the number of non-zero differences ≥20. The results of the Wilcoxon signed-rank test are summarised in Table 1. The mean bilirubin measured by the PoC system was M=97.74 μmol/L (SD=149.46 μmol/L), whilst the value measured by Alinity was M=103.35 μmol/L (SD=158.1 μmol/L). The results of the Wilcoxon signed-rank test found no statistically significant change using the PoC device and the standard laboratory method measurements, W=140, *P*=0.191, *n*=20. There were a total of 6 patients that reduced their serum bilirubin concentrations (negative rank sum=70) and 14 people that increased (positive rank sum=140).

Fig. SI3 illustrates the ROC plot. In this calculation, the clinically relevant threshold values for bilirubin concentrations of 200 μmol/L, 300 μmol/L and 450 μmol/L were used. The PoC measurement method demonstrates a sensitivity of 90% and specificity of 97% in categorising the samples within these bands. The area under the plot is over 0.82, corresponding to good detection capability. The *P*-value extracted from this plot is 0.000,4.

## 4. Discussions

Several PoC devices for the measurement of bilirubin concentration based on dual-wavelength operating principles have been reported in earlier works. Bilistick (Bilimetrix, Trieste, Italy), a PoC device, has a Pearson correlation coefficient of r=0.961 (*P*<0.001) against reference laboratory measurement. Moreover, it has a mean difference of +10.3 μmol/L and a 95% *CI* within the measurement range of −38 μmol/L and +58.7 μmol/L compared with the clinical laboratory reference measurement.[6] In another study, a mean difference of the Bilistick compared with reference laboratory measurement was reported as −15 μmol/L, with the 95% *CI* of −94 μmol/L to +63 μmol/L.[17] Moreover, a systematic review by Westenberg *et al.* (2023) showed an overall difference between the Bilistick and laboratory-based bilirubin measurement of −17 μmol/L, with a 95% level of accuracy range of −119 to +86 μmol/L.[18] BiliDx (UNISTAT Reichert Technologies) is another PoC device used for bilirubin measurement.[19] In recently reported work on BiliDx, the mean difference was reported as +0.75 mg/dL, and a 95% *CI* was −2.57 to +4.07 mg/dL.[20] These values correspond to a mean difference of 12.8 μmol/L, with a 95% *CI* of −43.7 μmol/L to +69 μmol/L.

Future studies with larger sample sizes are necessary for more robust conclusions. The results of the power analysis indicated that a minimum sample size of 176 participants would be necessary to adequately test for a significant change in clinical performance between the PoC device and the standard hospital device for bilirubin measurement. This analysis was based on an effect size of 0.21, an α level of 0.05, and a desired power of 0.80, with an expected mean difference of 1.7 μmol/L and a standard deviation of 8 μmol/L assessed in this pilot study. While 176 participants are the minimum required sample size to achieve the desired statistical power, it is imperative to consider that a larger sample may enhance the robustness of the findings. Without sufficient participants, the study may not adequately detect the hypothesized differences, potentially compromising the validity of the results and their implications for clinical practice.

Another key consideration is that prior studies are based on whole blood measurement, whilst in this work, we focused on blood serum. Although we have validated the setup for direct measurement of bilirubin in whole blood samples in our earlier studies using animal model samples,[12,13] further clinical studies must be performed to confirm this.

## 5. Conclusions

This study assessed the viability of using a dual wavelength based PoC system for bilirubin detection in a pilot cohort. The results presented here are based on a relatively small sample size of 20, and future larger-scale studies are required to further validate the clinical applicability of these findings. Nevertheless, results indicate a significant correlation between the bilirubin measured by the PoC system and the standard laboratory method, with a relatively small instrument bias. Additionally, analyses suggest that the PoC system offers high sensitivity and specificity, suitable for clinical decision-making over a broad range of bilirubin concentrations. Along with these key attributes, the PoC device is constructed using a miniaturised integrated sensing module with a microfluidic test cartridge and enables a simple and rapid measurement. Although earlier preclinical studies showed accurate detection in whole blood samples, method comparison studies should be performed using samples with a higher hemolysis index to assess the degree of hemolysis on correlation. Future research should also assess the correlation in different patient groups, such as neonates, adults with cirrhosis, well patients, and unwell patients.

## CRediT authorship contribution statement

**Jean Pierre Ndabakuranye:** Writing – review & editing, Writing – original draft, Investigation, Formal analysis, Data curation. **Inge W.G. Last:** Writing – original draft, Formal analysis. **Kay Weng Choy:** Writing – review & editing, Writing – original draft, Resources, Methodology, Investigation, Formal analysis, Data curation, Conceptualization. **Peter Thurgood:** Methodology. **Jason C. Steel:** Validation. **Genia Burchall:** Formal analysis. **Stella Stylianou:** Writing – review & editing, Formal analysis, Conceptualization. **Khashayar Khoshmanesh:** Writing – review & editing, Writing – original draft, Formal analysis, Data curation. **Arman Ahnood:** Writing – review & editing, Writing – original draft, Supervision, Methodology, Funding acquisition, Formal analysis, Conceptualization.

## Ethical approval

## Funding

This work was supported by a CASS Foundation (Medicine and Science) Grant and Australian Research Council through a Discovery Projects Grant (No. DP230100019). The authors also appreciate the access granted to the facilities and the scientific and technical assistance of the Australian Microscopy & Microanalysis Research Facility at RMIT University.

**Declaration of competing interest**

The authors declare that they have no known competing financial interests or personal relationships that could have appeared to influence the work reported in this paper.

**Acknowledgements**

The authors acknowledge the facilities, as well as the scientific and technical assistance of the RMIT University's Microscopy and Microanalysis Facility (RMMF), and the RMIT Micro Nano Research Facility (MNRF) in the Victorian Node of the Australian National Fabrication Facility (ANFF-Vic).

**Appendix A. Supplementary data**

Supplementary data to this article can be found online at https://doi.org/10.1016/j.lmd.2025.100073.